\documentclass[conference]{IEEEtran}%
\usepackage{graphicx,multicol,amsthm} 
\usepackage{epsfig} 
\usepackage{amsmath} 
\usepackage{amssymb}  
\usepackage{mathtools}
\usepackage{float}
\usepackage{cases,setspace,adjustbox,xspace}

\usepackage{texdef2011}
\newcommand{\nn}{\nonumber\\}

\usepackage{cite}

\ifCLASSINFOpdf
\else
\fi
\usepackage{algorithmic}
\usepackage[ruled,vlined]{algorithm2e}

\usepackage{array}

\usepackage[update,prepend]{epstopdf}

\usepackage{eqparbox}

\usepackage[font=footnotesize]{subfig}
\usepackage{fixltx2e}
\usepackage{url}

\newlength{\overwritelength}
\newlength{\minimumoverwritelength}
\newcommand{\overwrite}[3][red]{%
  \settowidth{\overwritelength}{$#2$}%
  \ifdim\overwritelength<\minimumoverwritelength%
    \setlength{\overwritelength}{\minimumoverwritelength}\fi%
  \stackrel
    {%
      \begin{minipage}{\overwritelength}%
        \color{#1}\centering\small #3\\%
        \rule{1pt}{9pt}%
      \end{minipage}}
    {\colorbox{#1!50}{\color{black}$\displaystyle#2$}}}

\newcommand{\Rmax}{R_{\max}}
\newcommand{\pmin}{p_{\min}}
\newcommand{\pmax}{p_{\max}}

\newcommand{\ageALOHA}{\age_{\text{ALOHA}}}
\newcommand{\ageSF}{\age_{\text{SF}}}

\newcommand{\age}{\Delta}
\newcommand{\maxSlot}{S}

\newcommand{\numNodes}{M}
\newcommand{\other}[1]{A_{#1}}
\newcommand{\residual}[1]{X_{#1}}

\newcommand{\Tavg}[2][t']{\left\langle{#2}\right\rangle_{#1}}

\newcommand{\SWF}{\emph{SF}\xspace}

\newcommand{\ALOHA}{\emph{ALOHA}\xspace}
\begin{document}
%
\title{\smaller Status Updates Over Unreliable Multiaccess Channels}
\author{\IEEEauthorblockN{Sanjit~K.~Kaul}
\IEEEauthorblockA{Wireless Systems Lab, IIIT-Delhi \\
E-mail : skkaul@iiitd.ac.in}
\and
\IEEEauthorblockN{Roy~D.~Yates}
\IEEEauthorblockA{WINLAB, ECE Dept., Rutgers University\\
Email: ryates@winlab.rutgers.edu}\thanks{This work was supported by NSF Award CIF-1422988 received by Roy Yates and the Young Faculty Research Fellowship (Visvesvaraya PHD scheme) received by Sanjit Kaul.}}
\IEEEoverridecommandlockouts



%


\newif\iflonger
\longertrue

\maketitle

\begin{abstract}
Applications like environmental sensing, and health and activity sensing, are supported by networks of devices (nodes) that send periodic packet transmissions over the wireless channel to a sink node. We look at simple abstractions that capture the following commonalities of such networks (a) the nodes send periodically sensed information that is temporal and must be delivered in a timely manner, (b) they share a multiple access channel and (c) channels between the nodes and the sink are unreliable (packets may be received in error) and differ in quality.

We consider scheduled access and slotted ALOHA-like random access. Under scheduled access, nodes take turns and get feedback on whether a transmitted packet was received successfully by the sink. During its turn, a node may transmit more than once to counter channel uncertainty. For slotted ALOHA-like access, each node attempts transmission in every slot with a certain probability. For these access mechanisms we derive the age of information (AoI), which is a timeliness metric, and arrive at conditions that optimize AoI at the sink. We also analyze the case of symmetric updating, in which updates from different nodes must have the same AoI. We show that ALOHA-like access, while simple, leads to AoI that is worse by a factor of about $2e$, in comparison to scheduled access.


\end{abstract}%

%
\IEEEpeerreviewmaketitle
\section{Introduction}
Applications across domains including healthcare (monitoring of patients), energy (smart meters and grids), buildings (appliance monitoring, temperature control), and environment monitoring (pollution information), will leverage temporal information obtained from large numbers of sensing devices.

The wireless access networks that will transport the sensed information from these devices to the Internet are expected to consist of hundreds of low power sensing devices (nodes) spread over a large area and connected to an access point, that provides connectivity to the Internet\footnote{An example effort towards the standardization of such networks is IEEE $802.11$ah/WiFi-HaLow:~\url{http://www.ieee802.org/11/Reports/tgah_update.htm}}. The nodes periodically send packets of sensed information (the node's \emph{state}) to the access point (\emph{sink}) over a multiaccess channel.

In this work we derive insights into the timely delivery of nodes' state to the sink using simple abstractions of such networks. We use the metric of age of information (AoI) to quantify \emph{timeliness}. Suppose node $i$'s state as known to the sink at time $t$ was current at time $u_i(t) < t$, then the \emph{age} of the node's state at time $t$ is the random process $\age_i(t) = t-u_i(t)$ and the AoI of the node is the average age.

We assume that the \emph{channel} between a node and the sink is unreliable and state update packets transmitted by the node are decoded in error and discarded by the sink with a certain non-zero probability. This probability may vary across nodes in the network. For the \emph{multiaccess} mechanism, we consider \emph{scheduled access with feedback} (\SWF) and \emph{slotted ALOHA-like random access} (\ALOHA).

In \SWF, nodes take turns sending their packets to the sink. The sink provides the nodes with instantaneous feedback on whether their packet transmission was decoded successfully. Each node is allowed up to a maximum of $S\ge 1$ packet transmission attempts during its turn. During its turn, a node transmits until its packet is decoded successfully or the allowed $\maxSlot$ attempts are exhausted. The selection of $\maxSlot$ is crucial to optimizing the AoI. Under \ALOHA, each node attempts transmission in every transmission slot with a certain probability. In \ALOHA, tuning of the attempt probability is essential for small AoI. 

Our contributions and the organization of the paper are as follows. Section~\ref{sec:related} summarizes  related work and Section~\ref{sec:network} details the network model. 
In Section~\ref{sec:schedAccessWithFeedback} we derive the AoI (Lemma~\thmref{ageSchedWithFeedback}) for the \SWF network. Our analysis leads to Lemma~\thmref{homogeneousNetWithFeedback} which states that if all nodes have the same success probability for packet decoding, then every node should be allowed to transmit during its turn until its transmission is successful. This section ends with observations for when nodes may have heterogeneous success probabilities.

We analyze \ALOHA in Section~\ref{sec:ALOHA}, where we present an approximation of nodes' attempt probabilities that minimize AoI. In Section~\ref{sec:symmetric}, we analyze \SWF and \ALOHA under the assumption of {\em symmetric updating}, in which all nodes must have the same AoI at the sink. We arrive at Theorem~\ref{thm:alohaVsSched}, which states that using \ALOHA, instead of \SWF, leads to AoI that is worse by a factor of approximately $2e$. We conclude the paper in Section~\ref{sec:conclusions}. 

\section{Related Work}
\label{sec:related}
Many recent works analyze the AoI under different system assumptions. We considered the M/M/1 first-come-first-served (FCFS) system with multiple sources in~\cite{2012ISIT-YatesKaul}. AoI for a single source sending updates when update packets may be delivered out-of-order has been analyzed by~\cite{KamKompellaEphremides2013ISIT,KamKompellaEphremides2014ISIT,Kam-PathDiversity2016}. The metric peak age of information (PAoI) was introduced in~\cite{CostaCodreanuEphremides2014ISIT} and has been studied by~\cite{HuangModiano2015ISIT} and~\cite{CostaCodreanuEphremides2016}. In~\cite{ChenHuang-ISIT2016}, the authors consider AoI in LCFS and FCFS M/M/1 systems with packet delivery error. 
Energy-constrained updating has been studied in\cite{Elif2015ITA,Yates2015ISIT,UpdateorWait-Infocom2016} in which updates are submitted to the server with knowledge of the server state.


In~\cite{kaul_minimizing_2011_short_cite}, we looked at minimizing the age of status updates sent by vehicles over a carrier-sense multiple access (CSMA) network. A local minimum was seen to exist in simulations. In~\cite{optimalLinkSchedICC2016}, the authors consider scheduling packets of $N$ sources, each of which is associated with a transmitter (TX) and a receiver (RX). The receivers are a hop away from their transmitters. Certain TX-RX links may be co-channel and cannot be scheduled together. Each source has a certain number of packets to send. The goal is to choose a schedule that minimizes the peak age of information in the network. In~\cite{broadcastModiano2016}, the authors consider the problem of a base station scheduling information to clients over a wireless network with unreliable channels. In our work we consider scheduled and random multiaccess. When the users are scheduled, our model for status updating is equivalent to that of wireless broadcasting considered in~\cite{broadcastModiano2016}.

\section{Multiaccess Network Model}
\label{sec:network}
\begin{figure}[t]
\centering
\includegraphics[scale=0.7]{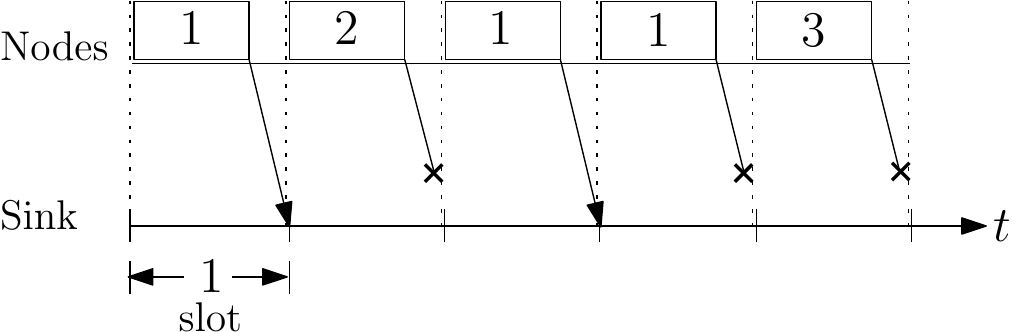}
\caption{\small{An illustration of the slotted system. The rectangles at the top of the figure denote the length of the packet transmission. They begin at the start of a slot. An arrow denotes the propagation of the packet. Each packet is labelled with the node sending it. An arrowhead is when the update is received by the sink, which happens at the end of the slot. A cross denotes a lost update.}}
\label{fig:slotted}
\end{figure}
We consider a network of nodes $1,2,\ldots,\numNodes$ that use a shared channel to transmit packets containing their current state to a sink. The goal of each node is to keep the sink \emph{updated} with its state. We assume a slotted system in which each node's packet transmission is a slot long. A transmission always starts at the beginning of a slot. A packet is received by the sink at the end of a slot and decoded instantaneously (we assume negligible packet processing time). Slot start and end times are synchronized across all nodes in the network. Figure~\ref{fig:slotted} illustrates a slotted system.

Let $p_i$ be the probability that a packet transmission by node $i$ is correctly decoded by the sink. A packet may be decoded incorrectly by the sink due to channel impairments. A correctly decoded packet updates the sink with node $i$'s state. We will refer to this event simply as an \emph{update} from node $i$. A packet that is not decoded correctly is \emph{lost}. The state of node $i$ stays unchanged at the sink. Packets decoded in error are not retransmitted. Every packet transmission is a new packet that contains the state of the transmitting node at the beginning of the packet transmission slot. We assume that $p_i$ is known for each node in the network. In practice, the same may be estimated for each node's link to the sink.

\begin{figure}[t]
\centering
\includegraphics[scale=0.54]{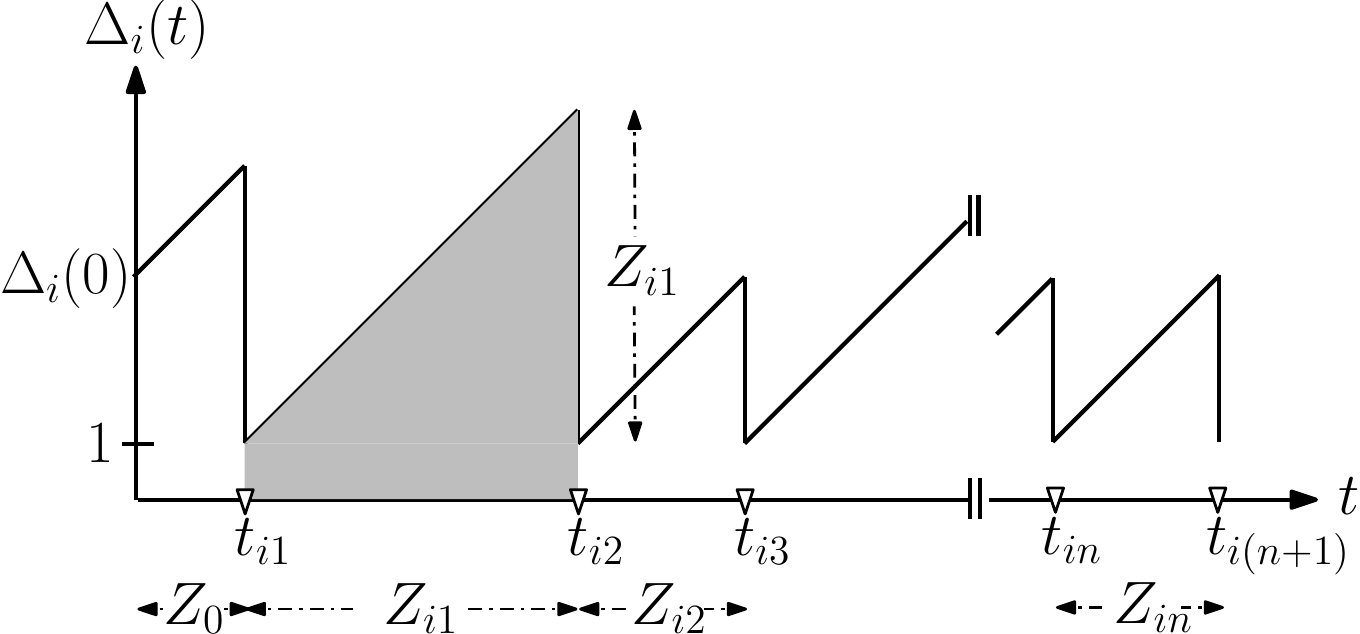}
\caption{\small{Example sample function of age $\age_i(t)$ of node $i$ at the sink. The time $t_{ij}$ is the time of the $j\textsuperscript{th}$ update from node $i$.}}
\label{fig:ageAccess}
\end{figure}

Let $t_{ij}$ be the time of the $j\textsuperscript{th}$ update by node $i$. 
The time between the $j\textsuperscript{th}$ and the $(j+1)\textsuperscript{th}$ update by node $i$ is the \emph{inter-update} time $Z_{ij} = t_{i(j+1)} - t_{ij}$. We will assume the $Z_{ij}$ of node $i$ are identically distributed as $Z_i$.

We will quantify the freshness of node $i$'s state at the sink in terms of the age of the state at the sink. Let the instantaneous age of node $i$'s state at the sink be given by $\age_i(t)$. Suppose that at time $t$ the sink has node $i$'s state that was current at time $t_0$. The age of node $i$'s state is $\age_i(t) = t-t_0$. Figure~\ref{fig:ageAccess} shows an example sample function of the instantaneous age $\age_i(t)$ of node $i$.  Age increases linearly with time between two updates, which is during intervals $Z_{ij}$ long, and is reset to $1$ slot at each $t_{ij}$ (because a packet transmission takes one slot).

Over an observation interval $(0,t')$, the time-average of the age function is
\begin{align}
\Tavg{\age_i} = \frac{1}{t'} \int_{0}^{t'} \age_i(t)\, dt.
\label{eqn:ageTimeAvg}
\end{align}

We define the age of information (AoI) $\Delta_i$ of node $i$ as
\begin{align}
\age_i = \lim_{t'\to \infty} \Tavg{\age_i}.
\label{eqn:ageOfInfi}
\end{align}

We will assume that this limit exists. Using arguments similar to those in~\cite{2012ISIT-YatesKaul}, also detailed in \iflonger Appendix~\ref{sec:derivationAoI}\else Appendix~A of~\cite{aloha-longer}\fi, we get
\begin{align}
\age_i = \frac{\E{Z_i^2}}{2\E{Z_i}} + 1.
\eqnlabel{avgAgei}
\end{align}
We define the AoI $\age$ of the network as
\begin{align}
\Delta = \frac{1}{M}\sum_{i=1}^{\numNodes} \Delta_i.
\label{eqn:avgAgeNetwork}
\end{align}
Next, we look at the \SWF and \ALOHA schemes.

\section{Scheduled Access With Feedback}
\label{sec:schedAccessWithFeedback}
In the \SWF multiaccess scheme, nodes take turns to transmit their state update packets. All nodes receive feedback from the sink about whether a transmitted packet was decoded correctly, at the end of the packet transmission. We assume that the overhead of the feedback is negligible in comparison to a transmission slot and will ignore it in the following analysis. 

During its turn, a node is allowed to transmit its packets a maximum of $\maxSlot$ times. A node's turn ends when it \emph{updates} the sink or when it has exhausted the allowed $\maxSlot$ transmission attempts. Thus a node's turn constitutes of a random number of one or more slots with a maximum value of $\maxSlot$. For such a multiaccess scheme, the value of $\maxSlot$ that minimizes the AoI $\age$ of the network, given by Equation~(\ref{eqn:avgAgeNetwork}), is of interest.

We know from~(\ref{eqn:avgAgei}) that the calculation of AoI $\Delta_i$ of node $i$ requires $\E{Z_i}$ and $\E{Z_i^2}$.  We now outline their derivation.
Consider the slot boundary at which an \emph{update} by node $i$ takes place. The random variable $Z_i$ is the time to the next update by the node. An update by node $i$, by assumption in this scheme, must also mark the end of node $i$'s turn. It must hence be followed by the turns of all other nodes before node $i$ gets its next turn. Since a packet transmission by node $i$ leads to an update with probability $p_i \le 1$, the next update by node $i$ may not take place for the next one or more of its turns.

\begin{figure}[t]
\centering
\includegraphics[scale=0.60]{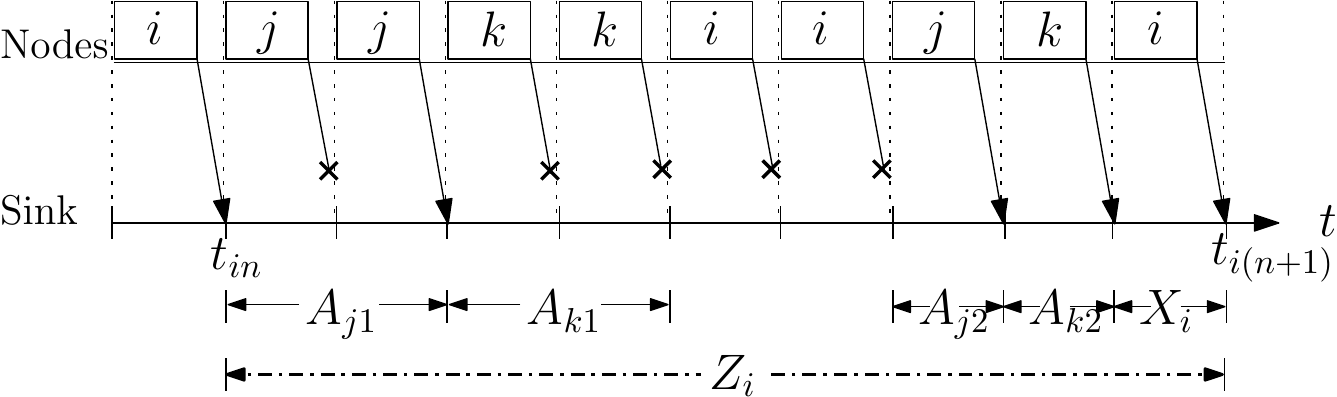}
\caption{\small{We have $\numNodes=3$ nodes using \SWF with $\maxSlot=2$ slots. The representation of packet transmissions is as in Figure~\ref{fig:slotted}. The interval $Z_i$ starts with the $n\textsuperscript{th}$ update by node $i$ at $t_{in}$. Next, node $j$ begins to transmits its state. The packet is lost and $j$ utilizes the second slot in its turn, at the end of which it updates. Thus, $\other{j1} = 2$. This is followed by node $k$ take $\other{k1} = 2$ slots. After this $i$ gets its next turn but both transmitted packets are lost. Then $j$ and $k$ take their turns. Both update at the end of the first slot in their respective turns ($\other{j2} = 1, \other{k2} = 1$). Finally, there is an update by node $i$ at the end of the first slot of its turn. We have $N_i=2$ and $\residual{i} = 1$.}}
\label{fig:schedWithFeedback}
\end{figure}

Let $N_i\ge 1$ be the number of turns taken by node $i$ to update the sink. Then node $i$ must have taken $N_i - 1$ turns, all $\maxSlot$ slots long, during which all its packet transmissions were \emph{lost} and it was unable to update. During its $N_i\textsuperscript{th}$ turn, however, an update by node $i$ must happen. Let $\residual{i}$ be the number of slots for which node $i$ transmits during its $N_i\textsuperscript{th}$ turn. Further, observe that each node $j\ne i$ also takes $N_i$ turns between updates by node $i$. Let $\other{jk}$, $1\le k\le N_i$, be the number of slots for which node $j$ transmits during its $k\textsuperscript{th}$ turn since the last update by node $i$. We can therefore write the inter-update time $Z_i$ as
\begin{align}
Z_i = (N_i - 1) \maxSlot + \sum_{\substack{j\ne i\\ 1\le j\le \numNodes}} \sum_{k=1}^{N_i} \other{jk} + \residual{i}.
\label{eqn:ZiSchedWithFeedback}
\end{align}
Figure~\ref{fig:schedWithFeedback} depicts $Z_i$ for $\numNodes = 3$ and $\maxSlot = 2$. Note that $Z_i$ in~(\ref{eqn:ZiSchedWithFeedback}) is a random sum of random variables. \iflonger In Appendix~\ref{sec:proofOfLemmaWithFeedback}, \else In Appendix B of~\cite{aloha-longer},\fi we detail the probability mass functions of the random variables that constitute $Z_i$ and summarize the steps to arrive at Lemma~\thmref{ageSchedWithFeedback} that we state next. Let $r_i= 1 - (1-p_i)^\maxSlot$ be the probability that node $i$ has at least one successful transmission in $\maxSlot$ slots. Let $\eta_{ji} = r_{j}/r_{i}$.
\begin{lemma}\thmlabel{ageSchedWithFeedback} 
For scheduled access with feedback, the first and second moments of the inter-update time of node $i$ are 
\begin{align}
&\E{Z_i} = \sum_{j=1}^\numNodes \frac{\eta_{ji}}{p_j},\\
&\E{Z_i^2} = \frac{2-p_i}{p_i^2} + \sum_{j\ne i} \biggl\{\frac{2}{p_j^2}\eta_{ji}^2 + \frac{2\maxSlot}{r_i p_j} (\eta_{ji} - 1)\nonumber\\ 
&+ \left(\frac{2-p_i}{p_ip_j} + \frac{2(1-r_j)}{p_j^2}\right)\eta_{ji}
+(2-r_i)\sum_{\substack{j'\ne j\\ j'\ne i}} \frac{\eta_{ji}\eta_{j^{'}i}}{p_j p_{j^{'}}}\biggr\}.
\end{align}
\end{lemma}
\subsection{Observations}
We start by considering a homogeneous network of nodes. Specifically, the probability with which a packet transmitted by a node is successfully decoded by the sink is $p_i = p$, for all $i$ and $0 < p < 1$. Clearly, in a homogeneous network, $r_i = r_j$ and $\eta_{ij} = 1$, for all $i,j$. 
The mean of the inter-update time $Z_i$, given by Lemma~\thmref{ageSchedWithFeedback}, becomes $\E{Z_i} = M/p$, for all nodes $i$. Note that this mean is independent of $\maxSlot$.

Now consider $\E{Z_i^2}$. It is easy to see that after setting $\eta_{ij} = 1$, for all $i,j$, minimizing $\E{Z_i^2}$ is equivalent to maximizing $r_i$, where $r_i \le 1$. This is achieved in the limit as $\maxSlot \to \infty$. This fact, together with a constant mean, implies that the AoI $\age_i$, in Equation~(\ref{eqn:avgAgei}), of any node $i$ in the network, and hence also the AoI of the network $\age$ (Equation~(\ref{eqn:avgAgeNetwork})), is minimized in the limit as $\maxSlot \to \infty$. We summarize this observation in Lemma~\thmref{homogeneousNetWithFeedback}.

\begin{lemma}\thmlabel{homogeneousNetWithFeedback}
For a homogeneous network of $M$ nodes that take turns to transmit their state to a sink and get feedback on whether an update of state occurred, the network's AoI is minimized by allowing a node to keep transmitting packets during its scheduled turn until an update by the node occurs. Specifically, it is minimized in the limit as $\maxSlot \to \infty$. 
\end{lemma}
Next we show via example that the above strategy is not AoI minimizing when nodes have different successful transmission probabilities.


\subsubsection*{Heterogeneous Network of Nodes} 
\begin{figure}[tb]
\centering
\includegraphics[width=3.5in]{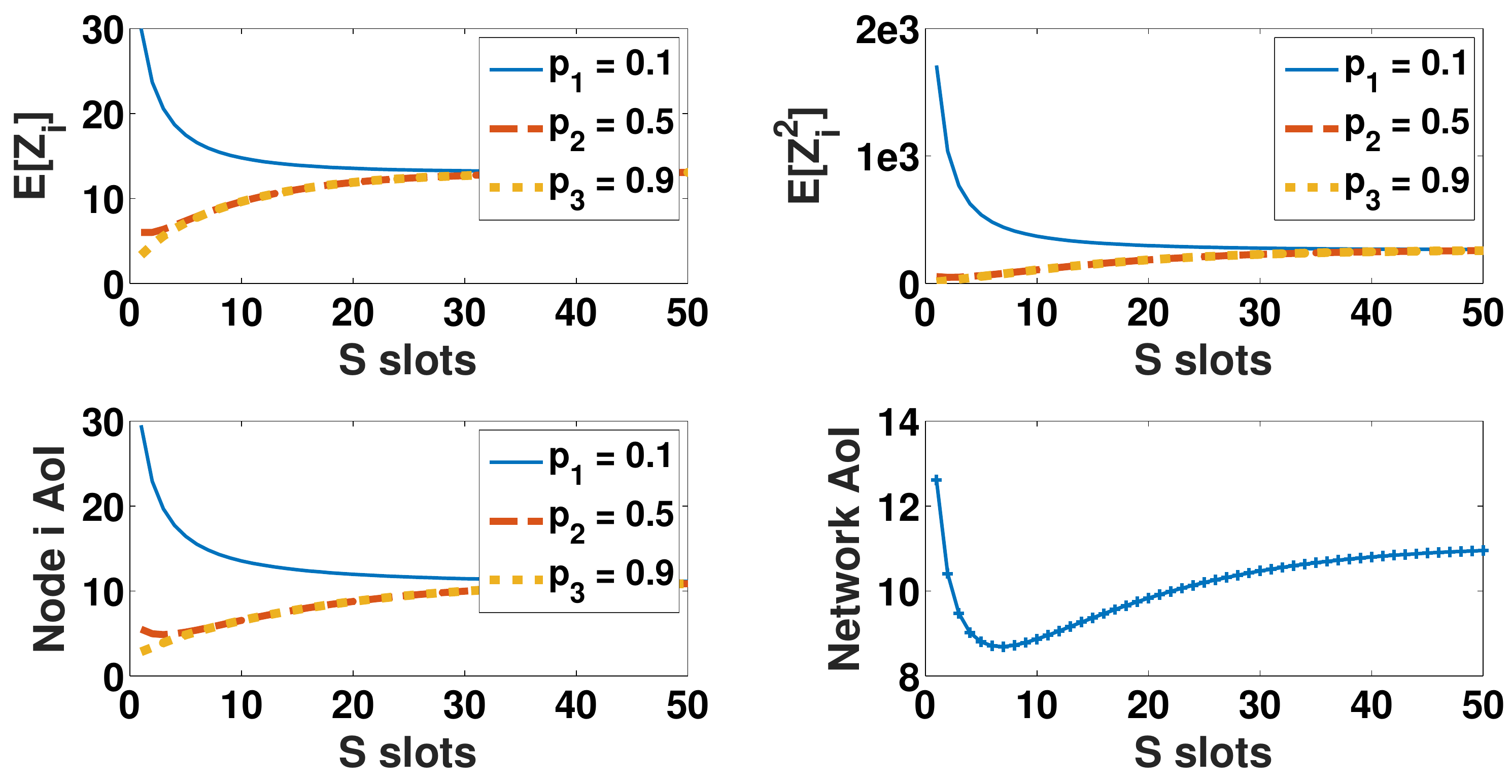}
\caption{\small{We show $\E{Z_i}$, $\E{Z_i^2}$, AoI $\age_i$ of node $i$, and the AoI $\age$ of a network of three nodes $1,2,3$ as a function of the maximum number of slots $\maxSlot$ in a turn. The probabilities of successful transmission are $p_1=0.1$, $p_2=0.5$, and $p_3=0.9$, respectively. Clearly, $S$ that is best for the AoI of a given node $i$, is not best for the network AoI.}}
\label{fig:hetNetExampleWithFeedback}
\end{figure}

We exemplify the behavior of AoI as a function of $\maxSlot$ in a heterogeneous network using a three node network. The nodes $1$, $2$, and $3$ have probabilities of successful transmission $p_1 = 0.1$, $p_2=0.5$, and $p_3 = 0.9$, respectively. In Figure~\ref{fig:hetNetExampleWithFeedback} we compare $\E{Z_i}$, $\E{Z_i^2}$, $\age_i$ for the three nodes and also show the AoI of the network as a function of the maximum number of slots $\maxSlot$ allowed in a turn. The AoI of the network is minimized at $S = 7$.

%

Intuitively, if the maximum number of slots is set to a relatively small value, a node with a poor channel (node $1$ in Figure~\ref{fig:hetNetExampleWithFeedback}) will often be interrupted by other nodes' turns, increasing the number of slots that elapse before a successful update and thus hurting the node's AoI. On the other hand, fixing the maximum number of allowed slots to a relatively large value will hurt a node with a very good channel as other nodes with poorer channels will end up using a large number of their allowed $S$ slots.

\section{Slotted Aloha-Like Random Access}
\label{sec:ALOHA}
In \ALOHA, a node $i$ attempts to transmit a packet  with probability $\tau_i$ in every slot. Unlike \SWF, no feedback is received from the sink. Our interest is finding the $\tau_i$, for all nodes $i$, that minimize the network AoI $\age$. If node $i$ is the only sender of state during a slot, then its packet is decoded correctly with probability $p_i$. If more that one packet transmissions occur during the same slot all the transmissions are decoded in error. Let $\gamma_i$ be the probability with which an update by node $i$ occurs at the end of a slot. The update occurs if node $i$ transmits during the slot, no other node transmits during the slot, and the node's packet is decoded correctly by the sink. Thus, we have
\begin{align}
\gamma_i = \tau_i p_i \prod_{j\ne i} (1-\tau_j).
\eqnlabel{probSuccUpdate}
\end{align}

The inter-update interval $Z_i$ is a geometric random variable with mean $\E{Z_i} = 1/\gamma_i$ and second moment $\E{Z_i^2} = 2/\gamma_i^2 - 1/\gamma_i$. Substituting in \eqnref{avgAgei}, node $i$ has age
\begin{equation}
\age_i =\frac{1}{2}+\frac{1}{\gamma_i}.
\end{equation}
From \eqnref{avgAgeNetwork}, the network AoI is
\begin{align}
\age = \frac{1}{2} + \frac{1}{\numNodes} \sum_{i=1}^\numNodes \frac{1}{\gamma_i}.
\label{eqn:ageAloha}
\end{align}

The first order optimality conditions are obtained by differentiating~(\ref{eqn:ageAloha}) with respect to $\tau_i$, $1\le i \le \numNodes$. They are
\begin{align}
\frac{1-\tau_i}{p_i\tau_i^2} = \sum_{j=1}^\numNodes \frac{1-\tau_j}{p_j\tau_j},\quad 1 \le i \le \numNodes.
\label{eqn:firstOrderCondExactAloha}
\end{align}
An exact closed form solution for the age minimizing $\tau_i = \tau_i^*$ is simple for a network of $\numNodes=2$ nodes. Solving~(\ref{eqn:firstOrderCondExactAloha}), we get $\tau_i^* = (1 + (p_i/p_j)^{(1/3)})^{-1}$, where $i,j\in \{1,2\}$ and $i\ne j$. We are unable to derive an exact closed form solution for $M>2$. An approximation for large $M$, when the optimal $\tau_i$ will be small, is given by
\begin{align}
\tau_i^* \approx \frac{(1/\sqrt{p_i})}{\sum_{j=1}^\numNodes (1/\sqrt{p_j})},\quad 1\le i\le M.
\end{align}
For the derivation of the approximation and an evaluation of its efficacy, we refer the reader to \iflonger Appendix~\ref{sec:appendixAloha}\else Appendix C of~\cite{aloha-longer}\fi.
\section{Symmetric Updating Systems}
\label{sec:symmetric}
When all nodes' state updates are equally important, it is desirable for all nodes to achieve the same AoI.  We refer to such updating systems as {\em symmetric}. We observe that \SWF and \ALOHA can both be configured to be symmetric systems. In this section, we compare them under symmetric operation.  While the decoding probabilities $p_i$ may be arbitrarily close to zero, symmetric updating requires $p_i>0$ for all nodes $i$; otherwise; the operation of the symmetric updating system will fail. In the following analysis, we assume $p_i\in[\pmin,\pmax]$ for all nodes $i$.

\SWF becomes symmetric as the maximum number of packet transmission attempts by a node, during a turn, $S\to\infty$. In its turn, node $i$ transmits packets in consecutive slots until it \emph{updates} the sink. Thus node $i$ transmits packets for $X_i$ time slots, where $X_i$ is a geometric $(p_i)$ random variable. Assuming the nodes are numbered in order of their turns, node $i$ is followed by node $i+1$ and so on until it's next turn. Node $i$ then has an inter-update time of
\begin{align}
Z_i=X_{i+1}+\cdots+X_{M}+X_1 +\cdots X_i
=\sum_{j=1}^MX_j.
\end{align}
Thus $Z_1,\ldots,Z_M$ are identically distributed and all nodes will obtain the same AoI. We can find the moments $\E{Z_i}$ and $\E{Z_i^2}$ either from first principles or from Lemma~\thmref{ageSchedWithFeedback} in the limit as $S\to\infty$ and thus  $r_i=1$ and $\eta_{ji}=1$. By either method, we obtain
\begin{align}
\E{Z_i}&=\sum_{j=1}^M \frac{1}{p_j},\\ 
\E{Z_i^2}&=(\E{Z_i})^2-\E{Z_i} +\sum_{j=1}^M\frac{1}{p^2_j}.
\end{align}
From \eqnref{avgAgei}, the average age of each node is 
\begin{align}\eqnlabel{ageSF}
\ageSF&=\frac{1}{2}\bracket{1+\sum_{j=1}^M\frac{1}{p_j}
+R(\pv)};\\
R(\pv)&=\frac{\sum_{j=1}^M\frac{1}{p_j^2}}{\sum_{j=1}^M\frac{1}{p_j}}.\eqnlabel{Rdefn}
\end{align}

We now turn to \ALOHA. The updating becomes symmetric when the success probabilities $\gamma_i$ of all nodes are the same. We observe that we can write \eqnref{probSuccUpdate} as
\begin{align}
\gamma_i = \frac{\tau_i p_i}{1-\tau_i} \prod_{j=1}^M (1-\tau_j).
\eqnlabel{probSuccUpdate2}
\end{align}
Since $\prod_{j=1}^M(1-\tau_j)$ is identical for all nodes, we can make the \ALOHA system symmetric by setting
\begin{align}\eqnlabel{beta_i}
\beta_i=\frac{p_i\tau_i}{1-\tau_i}=\beta,\quad 1\le i\le M.
\end{align}
In this case, $\beta$ is a parameter that can be tuned to minimize the AoI.  From \eqnref{beta_i}, node $i$ has attempt probability $\tau_i=\beta/(\beta+p_i)$ and from \eqnref{probSuccUpdate2} each node has update success probability
\begin{align}
\gamma(\beta) = \beta\prod_{j=1}^M \frac{p_j}{\beta+p_j}.
\eqnlabel{probSuccUpdate3}
\end{align}
Although $\gamma$ is not a concave function of $\beta$, the first order condition $d\ln(\gamma(\beta))/d\beta=0$ does yield a unique maximizer $\beta^*$ satisfying
\begin{align}
\eqnlabel{betastar}
 \sum_{j=1}^M\frac{\beta^*}{\beta^*+p_j}=1.
 \end{align}
 We observe that \eqnref{betastar} implies
 \begin{align}\eqnlabel{betabounds}
 \frac{\pmin}{M-1}\le \beta^*\le \frac{\pmax}{M-1}.
 \end{align}
 With $\beta=\beta^*$, each node has success probability $\gamma^*=\gamma(\beta^*)$ in each slot. Thus, for each node $i$, the inter-update time $Z_i$ is a geometric $(\gamma^*)$ random variable.
 It follows from \eqnref{avgAgei} that each node obtains AoI
 \begin{align}\eqnlabel{ageALOHA}
 \ageALOHA=\frac{1}{2}+\frac{1}{\gamma^*}.
 \end{align}
 We observe that the \SWF system requires substantial complexity in that the nodes must be ordered in a round-robin schedule and feedback is required to acknowledge successful transmissions. By contrast, the \ALOHA system is less demanding. Slot by slot feedback is not necessary and while the success probability can be optimized through $\beta$, this is not essential for operation. 
 
\begin{figure}[tb]
\centering
\includegraphics[width=3.0in]{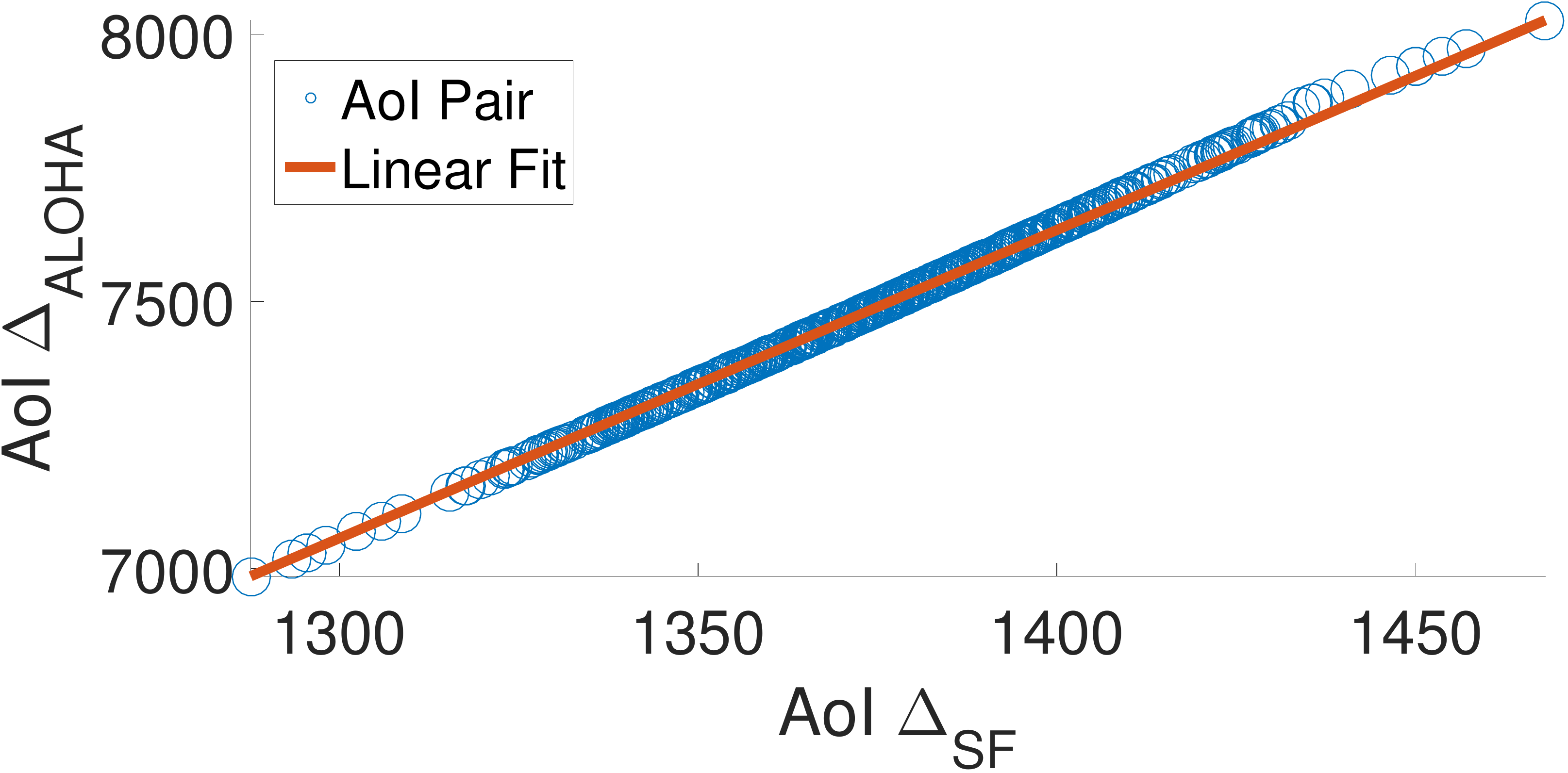}
\caption{\small{Scatter plot of ($\ageSF$, $\ageALOHA$) obtained from simulating $500$ networks of $1000$ nodes. For each network, the probability $p_i$ for each node $i$ was chosen uniformly and randomly from $(\pmin=0.1,\pmax=0.9)$. The fit has a slope $e^L=5.76$ (see~(\ref{eqn:Ldefn})) that is within the bound given by Theorem~\ref{thm:alohaVsSched}.}}
\label{fig:symUpdt}
\end{figure}

 Thus, the interesting open question is whether the scheduled access scheme is worth the additional complexity relative to the simpler \ALOHA system.  
The difficulty here is that it is hard to tell from  by directly comparing  \eqnref{ageSF} and \eqnref{ageALOHA}. With the definitions
\begin{align}\eqnlabel{Ldefn}
L&\equiv\ln\parfrac{\ageALOHA}{\ageSF}\quad\text{and}\quad \rho\equiv\frac{\pmax}{\pmin},
\end{align}
 the following theorem offers a simple comparison.
\begin{theorem}
\begin{align*}
\ln(2e)-L_M&\le L\le \ln(2e)+L_M\\
\shortintertext{where}
L_M&=\frac{1+2\rho^2}{M-1}+\frac{\rho^2}{(M-1)^2}.
\end{align*}
\label{thm:alohaVsSched}
\end{theorem}
Measured by AoI, the \SWF system is better by a factor of $2e\approx 5.4$ for large $M$. This is exemplified by the scatter plot in Figure~\ref{fig:symUpdt}. Intuitively, this result is easy to see. With a large number of nodes, the Aloha system has a packet success rate of $1/e$. Compared to the scheduled system, this increases the age by a factor of $e$. In addition, the Aloha system has geometric inter-update times, which roughly doubles the average age relative to scheduled round-robin inter-update times. For a proof of Theorem~\ref{thm:alohaVsSched}, we refer the reader to \iflonger Appendix~\ref{sec:proofAlohaVsSched}\else Appendix D of~\cite{aloha-longer}\fi.

\section{Conclusions}
\label{sec:conclusions}
We analyzed the age of information over multiaccess channels. For {scheduled access with feedback}, AoI for networks of nodes with similar probability of packet transmission error is optimized if the nodes are allowed to keep transmitting during their turn till a packet transmission is successful. For \emph{ALOHA-like random access}, we provided a closed form for the approximate transmission attempt probabilities that optimize AoI. We also considered symmetric updating systems and showed that for such systems, measured by AoI, using \ALOHA is worse than \SWF by a factor of about $2e$.
\appendices
\section{Derivation of Age of Information}
\label{sec:derivationAoI}
The time-average of the age function, over an observation interval $(0,t')$, is given by Equation~(\ref{eqn:ageTimeAvg}).

We assume that the age has a non-zero value $\age_i(0)$ at time $t=0$ when we begin observation. For simplicity of exposition, we will assume that the interval of observation ends at the time of the $(n+1)\textsuperscript{th}$ update. We have $t' = t_{i(n+1)}$. Using Figure~\ref{fig:ageAccess} to express the area under $\age_i(t)$ as a sum of the areas of the triangles of height and width $Z_{ij}$ and rectangles of height $1$ and width $T_{ij}$, we can rewrite the time-average age~(\ref{eqn:ageTimeAvg}) as
\begin{align}
\Tavg{\age_i} = \frac{T_0}{t'} + \frac{n}{t'} \frac{1}{n} \left[\sum_{j=1}^{n} \frac{Z_{ij}^2}{2} + Z_{ij}\right]\label{eqn:ageAsSumArea}.
\end{align}
where $T_0 = \frac{Z_0^2}{2} + \age_i(0) Z_0$.

We restate below the definition of the age of information (AoI) $\Delta_i$ of node $i$, earlier defined in Equation~(\ref{eqn:ageOfInfi}). It is given by
\begin{align}
\age_i = \lim_{t'\to \infty} \Tavg{\age_i}.
\end{align}
We will assume that this limit exists. 

Observe that in Equation~(\ref{eqn:ageAsSumArea}) $t' = t_{i(n+1)} = Z_0 + \sum_{j=1}^{n} Z_{ij}$ and $\lim_{t'\to \infty} T_0/t' = 0$. Also, as $t'\to\infty$, $n$ will become very large, and the sample averages on the right-hand-side of~(\ref{eqn:ageAsSumArea}) will converge to the corresponding statistical averages. Applying these observations to the AoI $\Delta_i$, given by Equation~(\ref{eqn:ageOfInfi}), yields the expression for $\Delta_i$ that is given in Equation~(\ref{eqn:avgAgei}).
\section{Proof of Lemma~\thmref{ageSchedWithFeedback}}
\label{sec:proofOfLemmaWithFeedback}
\begin{IEEEproof}
We continue the derivation of Lemma~\thmref{ageSchedWithFeedback} we had started in Section~\ref{sec:schedAccessWithFeedback}. Consider the random variable $N_i$. It can take integer values $\ge 1$. If $N_i = n$, then a node $i$ failed to update during $n-1$ turns, where each turn is of length $\maxSlot$ slots. Also, an update by the node occurred in one of the $\maxSlot$ slots during turn $n$. The event that a node $i$ is unable to update during a turn occurs with probability $(1-p_i)^\maxSlot$. Let $r_i = 1 - (1-p_i)^\maxSlot$. $N_i$ is a Geometric random variable with the probability mass function (PMF) given by
\begin{align}
P[N_i = n] = 
\begin{cases}
(1 - r_i)^{n-1} r_i & n\ge 1,\\
0 & \text{otherwise}.
\end{cases}
\label{eqn:pmfNiFeedback}
\end{align}
Consider the random variable $\other{jk}$. It is the number of slots in a turn taken by node $j$. Note that this number of slots can be $\maxSlot$ either when an update by node $j$ happens at the end of the last slot in the turn or when $j$ fails to update during the turn. When the number of slots is $a < \maxSlot$, the event is one in which the node fails to update in $(a-1)$ slots and the update occurs at the end of the slot $a$. Thus the PMF of $\other{jk}$ is given by
\begin{align}
P[\other{jk} = a] = 
\begin{cases}
(1-p_j)^{a-1} p_j& 1\le a < \maxSlot,\\
(1-p_j)^{a-1} & a = \maxSlot,\\
0 & \text{otherwise}.
\end{cases}
\label{eqn:otherFeedback}
\end{align}
Note that the $\other{jk}$ are identically distributed for all $k$. Let they be identically distributed as the random variable $\other{j}$. 

Consider the random variable $\residual{i}$. Note that $\residual{i}$ is the number of slots in a turn during which an update from node $i$ is known to have occurred. This conditioning on a guaranteed update gives us the PMF
\begin{align}
P[\residual{i} = x] = 
\begin{cases}
\frac{(1-p_i)^{x-1} p_i}{1 - (1-p_i)^\maxSlot} & 1\le x\le \maxSlot,\\
0 & \text{otherwise}.
\end{cases}
\label{eqn:pmfResidueFeedback}
\end{align}

Note that the random variables $\other{jk}$, $N_i$ and $\residual{i}$ are mutually independent. This fact and Equation~(\ref{eqn:ZiSchedWithFeedback}) allows us to write $\E{Z_i}$ and $\E{Z_i^2}$ in terms of the first and second moments of $\other{jk}$, $N_i$ and $\residual{i}$. We get
\begin{align}
\E{Z_i} &= \maxSlot (\E{N_i} - 1) + \E{N_i} \sum_{j\ne i} \E{\other{j}} + \E{\residual{i}},\label{eqn:EZiWithFeedback}\\
\E{Z_i^2} &= \sum_{\substack{j\ne i,\\1\le j \le \numNodes}} \biggl\{\E{N_i} \E{\other{j}^2} + \E{N_i^2 - N_i} (\E{\other{j}})^2\nonumber\\
&+ 2\maxSlot \E{N_i^2 - N_i}\E{\other{j}} + 2\E{\residual{i}} \E{N_i} \E{\other{j}}\nonumber\\
&+ \sum_{\substack{j' \ne j, j'\ne i,\\ 1\le j' \le \numNodes}} \E{N_i^2} \E{\other{j}} \E{\other{j'}}\biggr\} + \E{W^2},\label{eqn:EZi2WithFeedback}
\end{align}
where $W = (N_i - 1)\maxSlot + \residual{i}$.
The first and second moments of $N_i$, $\other{j}$, and $\residual{i}$ can be respectively obtained from the PMF(s) given by equations~(\ref{eqn:pmfNiFeedback}),~(\ref{eqn:otherFeedback}) and~(\ref{eqn:pmfResidueFeedback}). Substituting the moments into equations~(\ref{eqn:EZiWithFeedback}) and~(\ref{eqn:EZi2WithFeedback}) yields Lemma~\thmref{ageSchedWithFeedback}.
\end{IEEEproof}
\section{Attempt Probability Approximation for ALOHA}
\label{sec:appendixAloha}
\begin{figure}[tb]
\begin{center}
\includegraphics[scale=0.25]{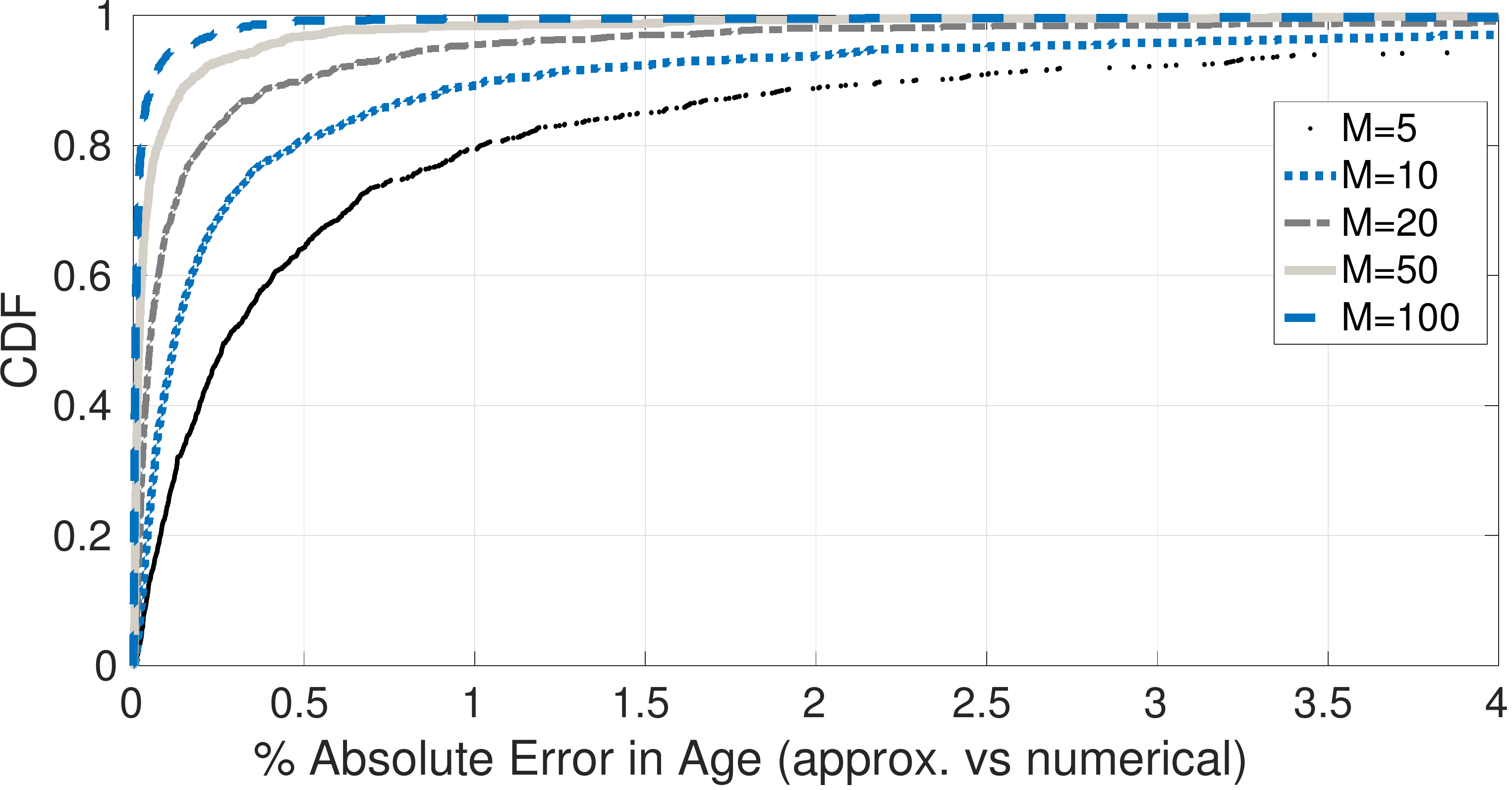}
\end{center}
\caption{\small{The cumulative distribution function (CDF) of the absolute percentage error in minimum age obtained using the approximation in~(\ref{eqn:approx4}) in comparison to that obtained from simulations that used a non-linear program solver to find the minimum age. The CDF is plotted for varied number of nodes $\numNodes$ in the network. Observe that even for $M=5$ about $95\%$ of the time the error is less than $5\%$. The approximation performs significantly better for larger $\numNodes$.}}
\label{fig:approxSimComp}
\end{figure}
It is reasonable to expect that for networks with large numbers of nodes the optimal attempt probability $\tau_i^*$ of node $i$ is small. To derive an approximate closed form solution, we will use the inequality $1 - x \le e^{-x}$ and that when $x$ is small $1 - x \approx e^{-x}$. Using the inequality and~(\ref{eqn:probSuccUpdate}), we can write
\begin{align}
\gamma_i \le p_i \frac{\tau_i}{1-\tau_i} e^{-\sum_{j=1}^{\numNodes} \tau_j}.
\end{align}
This and~(\ref{eqn:ageAloha}) gives us the lower bound $\hat{\age}$ on the age $\age$. We can write
\begin{align}
\age \ge \hat{\age} = \frac{1}{2} + \frac{e^{\sum_{j=1}^{\numNodes} \tau_j}}{\numNodes} \sum_{i=1}^\numNodes \frac{1}{p_i}\left(\frac{1}{\tau_i} - 1\right).
\end{align}
Note that we can approximate age $\age$ by $\hat{\age}$ for small $\tau_i$, when $1 - \tau_i \approx e^{-\tau_i}$. 
Define the quantities
\begin{align}
C = \sum_{j=1}^\numNodes \frac{1}{p_j}\left(\frac{1}{\tau_j} - 1\right)\quad\text{and}\quad C' = \sum_{j=1}^\numNodes \frac{1}{p_j\tau_j}.
\label{eqn:defineCCdash}
\end{align}

The first order optimality conditions for the approximate age $\hat{\age}$ are, for $1\le i\le \numNodes$,
\begin{align}
\frac{1}{p_i\tau_i^2} = C.\label{eqn:approx12}
\end{align}
From~(\ref{eqn:approx12}) we have $1/\tau_i = \sqrt{Cp_i}$ and from~(\ref{eqn:defineCCdash}) we know that $C < C'$. We can write
\begin{align}
\tau_i &\ge \frac{1}{p_i C' \tau_i} = \frac{\sqrt{C p_i}}{p_i C'} 
= \frac{\sqrt{C}}{\sqrt{p_i}} \frac{1}{\sum_{j=1}^\numNodes \frac{1}{p_j}\sqrt{C p_j}}\label{eqn:approx3}\\
&= \frac{(1/\sqrt{p_i})}{\sum_{j=1}^\numNodes (1/\sqrt{p_j})} := \hat{\tau}_i^{*} \approx \tau_i^{*}.\label{eqn:approx4}
\end{align}

The efficacy of the approximation $\hat{\tau}_i^{*}$ with respect to simulations carried out using a non-linear program solver is shown in Figure~\ref{fig:approxSimComp}. For each selection of $\numNodes$, we simulated $1000$ randomly selected vectors $[p_1,p_2,\ldots,p_\numNodes]$.


\section{Proof of Theorem~\ref{thm:alohaVsSched}}
\label{sec:proofAlohaVsSched}
\begin{IEEEproof} 
We first prove the lower bound. From \eqnref{ageALOHA}, 
\begin{align}
\ln\ageALOHA&\ge -\ln\gamma^*\nn
&=-\ln\beta^* -\sum_{j=1}^M\ln\parfrac{p_j}{p_j+\beta^*}\nn
&=-\ln\beta^* -\sum_{j=1}^M\ln\paren{1-\frac{\beta^*}{p_j+\beta^*}}\nn
&\ge-\ln\beta^* +\sum_{j=1}^M\frac{\beta^*}{p_j+\beta^*}
=-\ln\beta^*+1.\eqnlabel{alohaLB}
\end{align}
From \eqnref{Rdefn},
\begin{align}
R(\pv)\le \Rmax\equiv\frac{M/\pmin^2}{M/\pmax}=\frac{\pmax}{\pmin^2}.\eqnlabel{Rmaxdefn}
\end{align}
It then follows from \eqnref{ageSF}, \eqnref{Ldefn} and \eqnref{alohaLB} that
\begin{align}
L&\ge -\ln\beta^*+1+\ln2
-\ln\Bigr[1+\sum_{j=1}^M\frac{1}{p_j}+\Rmax\Bigl]\nn
&=\ln(2e)-\ln\Bigl[\beta^*(1+\Rmax)+\sum_{j=1}^M\frac{\beta^*}{p_j}\Bigr].\eqnlabel{LLB}
\end{align}
Now we observe that
\begin{align}
\sum_{j=1}^M\frac{\beta^*}{p_j}
&=\beta^*\sum_{j=1}^M\bracket{\frac{1}{p_j+\beta^*}
+\frac{\beta^*}{p_j(p_j+\beta^*)}}\\
&=1+\sum_{j=1}^M\frac{(\beta^*)^2}{p_j(p_j+\beta^*)}\eqnlabel{betasum2}\\
&\le 1+\parfrac{\pmax}{M-1}^2\frac{M}{\pmin^2}
=1+\frac{M\rho^2}{(M-1)^2}.\eqnlabel{betasum3}
\end{align}
Note that \eqnref{betasum2} follows from \eqnref{betastar} and \eqnref{betasum3} is a consequence of \eqnref{betabounds}. Applying \eqnref{Rmaxdefn} and \eqnref{betasum3} to \eqnref{LLB} yields
\begin{align}
L&\ge\ln(2e)-\ln\bracket{1+\frac{\pmax(1+\Rmax)}{M-1}
+\frac{M\rho^2}{(M-1)^2}}\nn
&\ge\ln(2e)-\frac{\pmax+\rho^2}{M-1}
-\frac{M\rho^2}{(M-1)^2}.
\end{align}
The lower bound follows from $\pmax\le1$ and some algebra.
For the upper bound, we observe that \eqnref{ageSF} and \eqnref{ageALOHA} imply
\begin{align}
\ln\ageSF&\ge -\ln2+\ln\Bigr[1+\sum_{j=1}^M\frac{1}{p_j}\Bigl],\\
\ln\ageALOHA&\le \ln(1+1/\gamma^*).
\end{align}
This implies
\begin{align}
L\le \ln2+\ln\bracket{\frac{1+1/\gamma^*}{1+\sum_{j=1}^M 1/p_j}}.
\end{align}
Using the inequality 
\begin{align}
\ln\frac{1+x}{1+y}\le \ln\frac{x}{y}=\ln x-\ln y,
\end{align}
we obtain
\begin{align}
L&\le \ln2+\ln\frac{1/\gamma^*}{\sum_{j=1}^M \frac{1}{p_j}}
= \ln2-\ln\gamma^*-\ln\sum_{j=1}^M\frac{1}{p_j}.\eqnlabel{LUB1}
\end{align}
Now we observe from \eqnref{ageALOHA} that
\begin{align}
-\ln\gamma^*&=-\ln\beta^*+\sum_{j=1}^M\ln\paren{1+\frac{\beta^*}{p_j}}\nn
&\le -\ln\beta^*+\sum_{j=1}^M \frac{\beta^*}{p_j}.
\end{align}
It then follows from \eqnref{betasum3} that
\begin{align}\eqnlabel{LGLB1}
-\ln\gamma^*\le -\ln\beta^*+1+\frac{M\rho^2}{(M-1)^2}.
\end{align}
Combining \eqnref{LUB1} and \eqnref{LGLB1} yields
\begin{align}\eqnlabel{LUB2}
L\le \ln(2e)+\frac{M\rho^2}{(M-1)^2}-\ln\paren{\sum_{j=1}^M\frac{\beta^*}{p_j}}.
\end{align}
It follows from \eqnref{betabounds} that $\sum_{j=1}^M\beta^*/p_j\ge 1$ and thus $\ln(\sum_{j=1}^M\beta^*/p_j)\ge 0$. Hence, the last term in \eqnref{LUB2} can be discarded. The upper bound then follows.
\end{IEEEproof}

\begin{spacing}{1.0}
\bibliographystyle{IEEEtran}
\bibliography{paper,ry-it}

\begin{thebibliography}{10}
\providecommand{\url}[1]{#1}
\csname url@samestyle\endcsname
\providecommand{\newblock}{\relax}
\providecommand{\bibinfo}[2]{#2}
\providecommand{\BIBentrySTDinterwordspacing}{\spaceskip=0pt\relax}
\providecommand{\BIBentryALTinterwordstretchfactor}{4}
\providecommand{\BIBentryALTinterwordspacing}{\spaceskip=\fontdimen2\font plus
\BIBentryALTinterwordstretchfactor\fontdimen3\font minus
  \fontdimen4\font\relax}
\providecommand{\BIBforeignlanguage}[2]{{%
\expandafter\ifx\csname l@#1\endcsname\relax
\typeout{** WARNING: IEEEtran.bst: No hyphenation pattern has been}%
\typeout{** loaded for the language `#1'. Using the pattern for}%
\typeout{** the default language instead.}%
\else
\language=\csname l@#1\endcsname
\fi
#2}}
\providecommand{\BIBdecl}{\relax}
\BIBdecl

\bibitem{2012ISIT-YatesKaul}
R.~Yates and S.~Kaul, ``Real-time status updating: Multiple sources,'' in
  \emph{Proc.~IEEE Int'l.~Symp.~Info.~Theory}, Jul. 2012.

\bibitem{KamKompellaEphremides2013ISIT}
C.~Kam, S.~Kompella, and A.~Ephremides, ``Age of information under random
  updates,'' in \emph{Proc.~IEEE Int'l.~Symp.~Info.~Theory}, 2013, pp. 66--70.

\bibitem{KamKompellaEphremides2014ISIT}
------, ``Effect of message transmission diversity on status age,'' in
  \emph{Proc.~IEEE Int'l.~Symp.~Info.~Theory}, June 2014, pp. 2411--2415.

\bibitem{Kam-PathDiversity2016}
C.~Kam, S.~Kompella, G.~D. Nguyen, and A.~Ephremides, ``Effect of message
  transmission path diversity on status age,'' \emph{IEEE Trans.~Info Theory},
  vol.~62, no.~3, pp. 1360--1374, March 2016.

\bibitem{CostaCodreanuEphremides2014ISIT}
M.~Costa, M.~Codreanu, and A.~Ephremides, ``Age of information with packet
  management,'' in \emph{Proc.~IEEE Int'l.~Symp.~Info.~Theory}, June 2014, pp.
  1583--1587.

\bibitem{HuangModiano2015ISIT}
L.~Huang and E.~Modiano, ``Optimizing age-of-information in a multi-class
  queueing system,'' in \emph{Proc.~IEEE Int'l.~Symp.~Info.~Theory}, Jun. 2015.

\bibitem{CostaCodreanuEphremides2016}
M.~Costa, M.~Codreanu, and A.~Ephremides, ``On the age of information in status
  update systems with packet management,'' \emph{IEEE Trans.~Info Theory},
  vol.~62, no.~4, pp. 1897--1910, April 2016.

\bibitem{ChenHuang-ISIT2016}
K.~Chen and L.~Huang, ``Age-of-information in the presence of error,'' in
  \emph{Proc.~IEEE Int'l.~Symp.~Info.~Theory}, 2016, pp. 2579--2584.

\bibitem{Elif2015ITA}
B.~T. Bacinoglu, E.~T. Ceran, and E.~Uysal-Biyikoglu, ``Age of information
  under energy replenishment constraints,'' in \emph{Proc.\ Info.\ Theory and
  Appl. (ITA) Workshop}, Feb. 2015, la Jolla, CA.

\bibitem{Yates2015ISIT}
R.~Yates, ``Lazy is timely: Status updates by an energy harvesting source,'' in
  \emph{Proc.~IEEE Int'l.~Symp.~Info.~Theory}, 2015.

\bibitem{UpdateorWait-Infocom2016}
Y.~Sun, E.~Uysal-Biyikoglu, R.~Yates, C.~E. Koksal, and N.~B. Shroff, ``Update
  or wait: How to keep your data fresh,'' in \emph{IEEE INFOCOM 2016 - The 35th
  Annual IEEE International Conference on Computer Communications}, April 2016,
  pp. 1--9.

\bibitem{kaul_minimizing_2011_short_cite}
S.~Kaul, M.~Gruteser, V.~Rai, and J.~Kenney, ``Minimizing age of information in
  vehicular networks,'' in \emph{{IEEE} {(SECON)}}, Salt Lake City, Utah,
  {USA}, 2011.

\bibitem{optimalLinkSchedICC2016}
Q.~He, D.~Yuan, and A.~Ephremides, ``On optimal link scheduling with min-max
  peak age of information in wireless systems,'' in \emph{2016 IEEE
  International Conference on Communications (ICC)}, May 2016.

\bibitem{broadcastModiano2016}
I.~Kadota, E.~Uysal-Biyikoglu, R.~Singh, and E.~Modiano, ``Minimizing the age
  of information in broadcast wireless networks,'' in \emph{2016 54th Annual
  Allerton Conference on Communication, Control, and Computing (Allerton)},
  Sept 2016, pp. 844--851.

\end{thebibliography}
\end{spacing}

%
%

\end{document}